**Title:** GraphVar 2.0: a user-friendly toolbox for machine learning on functional connectivity measures

**Authors:** Waller,. L.[1]*, Brovkin, A.[1,2]*, Dorfschmidt, L.[1,2], Bzdok, D.[3, 4, 5], Walter, H.[1], Kruschwitz, J.D. [1,2]

(*equal contribution)

**Affiliations:**

[1] Charité – Universitätsmedizin Berlin, corporate member of Freie Universität Berlin, Humboldt-Universität zu Berlin, and Berlin Institute of Health, Department of Psychiatry and Psychotherapy, Division of Mind and Brain Research, Germany

[2] Collaborative Research Centre (SFB 940) "Volition and Cognitive Control", Technische Universität, Dresden, Germany

[3] Department of Psychiatry, Psychotherapy and Psychosomatics, RWTH, Aachen University, 52072 Aachen, Germany

[4] JARA BRAIN, Jülich-Aachen Research Alliance, Germany

[5] Parietal team, INRIA, Neurospin, bat 145, CEA Saclay, 91191 Gif- sur- Yvette, France

∗ Corresponding author: Johann D. Kruschwitz (e-mail: johann.kruschwitz@charite.de), Division of Mind and Brain Research, Department of Psychiatry and Psychotherapy, Charitéplatz 1, 10117 Berlin, Germany

**Abstract:**

**Background:**
Background: We previously presented GraphVar as a user-friendly MATLAB toolbox for comprehensive graph analyses of functional brain connectivity. Here we introduce a comprehensive extension of the toolbox allowing users to seamlessly explore easily customizable decoding models across functional connectivity measures as well as additional features.

**New Method:** GraphVar 2.0 provides machine learning (ML) model construction, validation and exploration. Machine learning can be performed across any combination of graph measures and additional variables, allowing for a flexibility in neuroimaging applications.

**Results:** In addition to previously integrated functionalities, such as network construction and graph-theoretical analyses of brain connectivity with a high-speed general linear model (GLM), users can now perform customizable ML across connectivity matrices, graph measures and additionally imported variables. The new extension also provides parametric and nonparametric testing of classifier and regressor performance, data export, figure generation and high quality export.

**Comparison with existing methods:** Compared to other existing toolboxes, GraphVar 2.0 offers (1) comprehensive customization, (2) an all-in-one user friendly interface, (3) customizable model design and manual hyperparameter entry, (4) interactive results exploration and data export, (5) automated queue system for modelling multiple outcome variables within the same session, (6) an easy to follow introductory review.

**Conclusions:** GraphVar 2.0 allows comprehensive, user-friendly exploration of encoding (GLM) and decoding (ML) modelling approaches on functional connectivity measures making big data neuroscience readily accessible to a broader audience of neuroimaging investigators.

# 1.0 Introduction

## 1.1 Background

The increasing popularity of non-invasive neuroimaging techniques gave rise to vast amounts of data capturing the structural and functional architecture of the human brain (Van Essen et al. 2012 ; Van Horn et Toga 2014; Thompson et al. 2014; Eickhoff et al. 2016; Miller et al. 2016). The neuroimaging community thus faces a growing need for comprehensive methods and reliable tools (Poldrack et al. 2017; Smith et Nichols 2018) to derive rigorous neuroscientific conclusions from such datasets.

Traditional statistical approaches, including the general linear model (GLM), the workhorse of neuroimaging analysis, attempt to localize neural activity that correlates with behavior teased apart through careful hypothesis driven experimental design (Bzdok 2017a; Jack et al. 2018). The inferential conclusions drawn from such investigations, however, have repeatedly faced criticism over limitations, i.e. reproducibility (Bennett et Miller 2010; Pashler et Wagenmakers 2012). Given the accelerating availability of data, a purely hypothesis driven approach to cognitive investigations may no longer suffice. The application of machine learning, i.e. decoding, models allows for a complementary, pattern-oriented exploration of neuroimaging data, offering a promising solution to the identification of previously overlooked mechanisms. To this end, machine learning models employing functional connectivity measures hold promise in the search for a more informed stratification of clinical groups (Guo et al. 2012; Drysdale et al. 2017; Hojjati et al. 2017).

Brain network theory, or connectomics is steadily gaining momentum in the evaluation of the whole brain as an interconnected network, emerging as a powerful tool for capturing the complexity of the brains function and structure (Bullmore et Bassett 2011; Gu et al. 2015; Sporns et Betzel 2016). Due to ease of acquisition and a widespread prevalence of magnetic resonance imaging (MRI) facilities in the clinical setting, resting-state functional magnetic resonance imaging (rs-fMRI) has emerged as a promising non-invasive neuroinvestigative tool, measuring spontaneous fluctuations in blood oxygen level dependent (BOLD) signal at rest that reflect baseline neuronal activity. Aspects of rs-fMRI analysis still present ongoing methodological challenges (Cole et al. 2010; Murphy et al. 2013; Power et al. 2015; Bright et Murphy 2015). However, rs-fMRIs low subject compliance requirements, the convenient acquisition of complementary within-session structural MRI data, coupled with fMRIs high spatial resolution, continue to support its popularity in the investigation of intrinsic functional connectivity. Such investigations persist to complement inferences gained from other non-invasive neuroimaging modalities such as magnetoencephalography (MEG) and electroencephalography (EEG).

**1.2 Drawing conclusions from data: Encoding and decoding models**

As interpretation from data may not always be intuitive, it is important to distinguish between encoding models (e.g., general linear models) and decoding models (e.g. machine learning models) when making inference from data in neuroimaging investigations (Naselaris et al. 2010; Haufe et al. 2014). Researchers may choose between the two complementary approaches, guided by whether they are trying to determine the origin of a neural process through hypothesis driven analysis or aim to extract information from the data. Both these flavors of statistical approaches are available in GraphVar.

Encoding, or forward, models express the observed data as functions of some underlying variables. They explain or model the generation of observed data, which is why they are often referred to as "generative" models. When building a forward model, (i.e. with GLM) one would ask: "How well does a variable or set of variables explain the generation of the observed data?". Generative models are thus useful in hypothesis driven research. GLM employs normality or Gaussian assumptions and an explicit error distribution model. Going a step further, one may choose to examine such findings in the context of existing frameworks and experimental insights.

Decoding, or backward, models on the other hand, extract latent factors as functions of observed data, thereby attempting to reverse the data generating process. This is different from a generative model, since one is concerned with extracting hidden processes that may contribute to the observed data free from a bias of possible generative mechanisms. In the search for biomarkers, model data may include other variables in addition to neuroimaging measures (i.e. genetic, behavioral descriptors). When building a backward model (i.e., with ML) one is asking: "Which measures contribute to the observed data?". Going a step further, one would examine the contribution or significance of individual features to the observed data.

This notion of data-driven exploration, deploying advanced pattern-learning algorithms with the aim of advancing tailored healthcare and automated medical decision making, is currently trending in precision psychiatry (Bzdok 2017b) with efforts to extract informative features from large datasets including neuroimaging measures, both on a within subject and group level, providing a promise towards progress in redefining traditional stratification of clinical groups (Lord et al. 2012; Drysdale et al. 2017; cf. Shen et al. 2017). Decoding models, thus hold great promise as tools for extrapolating potentially informative features from traditionally feature-rich neuroimaging datasets. Decoding and encoding models can be used to complement each other in the search for insights into cognitive mechanisms in health and disease.

**1.3 Large-scale network analysis and big data – Need for comprehensive tools**

Traditionally programming expertise and in-depth technical knowledge is required to overcome the challenges posed by functional connectivity analysis. Common challenges include 1) the construction of an appropriate analysis pipeline, 2) the reproducibility of such a pipeline once established in terms of connectivity analysis as well as appropriate statistical evaluation, 3) a lack of user friendly design allowing for easy customization of pipelines for the inexperienced user, 4) shortcomings in the visualization and interpretability of results. To tackle these current shortcomings of the neuroimaging toolset, the GraphVar toolbox (Kruschwitz et al. 2015) emerged as a GUI-based tool covering the growing need for a comprehensive, user friendly exploration of functional brain connectivity as well as the appropriate encoding model construction and evaluation associated with such analyses.

Since its release, GraphVar has been downloaded more than 4900 and cited 40 times at the time of writing, and has been used across resting state as well as task-based investigations (e.g., Vatansever et al. 2015; Polli et al. 2016; Voss et al. 2016; Golbabaei et al. 2016; Caminiti et al. 2016; Flodin et al. 2017; Hojjati et al. 2017; Bolt et al. 2017; Wu et al. 2017; Sala et al. 2017).

Several toolboxes are currently available for the study of brain connectivity, including the Brain Connectivity Toolbox (Rubinov et Sporns 2010), eConnectome (He et al. 2011), GAT (Hosseini et al. 2012), CONN (Whitfield-Gabrieli et Nieto-Castanon 2012), BrainNet Viewer (Xia et al. 2013), GTG (Spielberg 2014), BASCO (Göttlich et al. 2015), GRETNA (Wang et al. 2015), BRAPH (Mijalkov et al. 2017), with some toolboxes such as DynamicBC (Liao et al. 2014) and BSMART (Cui et al. 2008) allowing for dynamic connectivity analyses. However, to date, none of these toolboxes seem to offer the customizable construction of decoding models across a comprehensive selection of functional connectivity measures. On top of presenting a solution to the commonly reported challenges of functional brain connectivity analysis, the new GraphVar ML extension allows the user to build, validate and examine network based decoding models. GraphVar ML was developed under the GNU General Public License v3.0, and is thus fully open source and extensible by experienced users. All GraphVar 2.0 functionalities are available on MATLAB versions 8.4 (2014b) and upwards. As such, GraphVar 2.0 is both UNIX and Windows OS compatible. Some functionalities rely on the Statistics and Machine Learning Toolbox. Additionally, the Parallel Computing Toolbox is required if multiple parallel workers are selected to speed up performance.

## 2.0 Methods: Model Construction and Validation

The aim of the subsequent sections is to provide a practical overview of core concepts with regard to building and assessing decoding models that may involve graph theoretic metrics. Notably, GraphVar ML decoding models may include various feature types. Here we justify the methods implemented following the logical analysis workflow (Fig. 1) involved in constructing, training and evaluating a ML model in GraphVar.

## 2.1 Building a prediction model – GraphVar ML Interface

The new GraphVar ML extension allows users to build prediction models including measures calculated using the user-friendly GUI as well as any additional measures imported from a variable sheet. If desired, the user may choose dynamic instead of static functional connectivity measures, generated with the sliding window approach (Hutchinson et al. 2013). To build a model, users should then choose an outcome variable, i.e. prediction target, for their model.

The interface also allows users to run an additional model on a nuisance variable in order to assess its contribution to the model (Rao et al. 2015). Users can assess the influence of nuisance variables such as age and gender on the predictive value of the model, which may promote more insightful conclusions on the validity of models. Depending on the outcome variable type (i.e. categorical or continuous), the user may select between building a classification or regression model. Support Vector (SV) or Elastic Net model learning approaches are possible. GraphVar ML allows users to customize the model validation algorithm using a standard or nested K-fold cross-validation design. The nested cross-validation option offers a 3-step nested cross-validation structure (Whelan 2014) with in-built model selection (Fig. 4).

Finally, similar to the GLM, model performance may be evaluated by formal assessments of significance as enabled by parametric null-hypothesis testing or non-parametric permutation testing (Nichols et Holmes 2002). An interactive results viewer allows easy interpretation and evaluation of prediction results including various ML metrics, as well as parametric and non-parametric p-values. For convenience GraphVar also offers the analysis of multiple outcome variables, results and plot export.

**2.2 Selecting a Prediction Target and Features**

Multiple prediction targets may be selected and will be pipelined to be executed consecutively, appearing stacked in the results viewer. GraphVar ML allows the user to select a multitude of features to populate the design matrix (Fig.3). Users may choose between graph theoretical measures, raw connectivity matrix and additionally imported variables from a spreadsheet.

Machine learning models typically require that the features that make up the design matrix are pre-processed prior to model training. Functional connectivity measures calculated with the GraphVar GUI do not need additional pre-processing prior to model construction assuming that adequate preprocessing was performed prior to timeseries or connectivity matrix import. As mentioned in the introduction, rs-fMRI data is rich in confounds and noise, and thus requires additional preprocessing where the choice of preprocessing steps may greatly impact reproducibility (c.f. Murphy et al. 2013, Bright et Murphy 2015, Parkes et al 2017). While some preprocessing tools are currently available, there is currently no gold standard on how to definitively deal with these issues (c.f. Yan et al. 2016, Esteban et al. 2018). Please refer to Appendix II, where a short summary about the currently existing literature on the issue of critical rs-fMRI preprocessing and network construction choice is provided.

Any externally imported additional variables which were not generated using the GraphVar pipeline should also be quality controlled for continuity, errors in entry such as duplication, noise and extreme outliers if they are added as features to the design matrix (Witten et al. 2016). Once the selected features are pooled into a design matrix, feature scaling is implemented in GraphVar ML to ensure standardization via Z-score normalization. Z-score normalization rescales features to ensure a standard normal distribution with a mean of 0 and a standard deviation of 1.

**2.3 Regularization Method**

The application of many ML models arises in data-settings where an abundance of input variables are available from each participant. The flipside of such data richness, is that classification or regression algorithm may falsely extrapolate patterns of statistical regularity. Often used to correct for overfitting, model regularization can be seen as a required step towards improving the performance of a predictive model (Bühlmann et Hothorn 2007). Regularization adds a penalty on the different parameters of the model, reducing its freedom. Consequently, the model will be less likely to fit the noise of the training data and which in turn improves the generalization abilities of the model to new data. GraphVar ML currently supports the use of Support Vector (SV) (Cortes et Vapnik 1995) and Elastic Net (Zou et Hastie 2005) learning methods. The choice between an Elastic Net or SV based algorithm requires an understanding of the difference between L1 and L2 regularization. Linear learning models may be penalized using an L1 or L2 regularization approach.

L1 allows feature selection, promoting sparsity. Sparse models are traditionally easier to interpret, with zero weight features deemed unimportant for prediction. When dealing with correlated features, L1 regularization will select only one out of a set of correlated features in a winner-takes-all approach. However, if the winning feature does not generalize well, model predictions may be less robust. L2, on the hand, does not offer feature selection and therefore does not promote sparsity. Elastic Net employs a combination of both. SV can in theory be trained using either L1 or L2 regularization, but our current implementation uses neither. Instead, it relies only on the power of separating classes by a hyperplane with a variable margin for generalization. Because of its inbuilt trade-off, Elastic Net may be seen as advantageous over SV, however SV may in many cases be computationally less expensive and suitable for larger datasets.

Within GraphVar ML, the SV algorithm is implemented using the established LIBSVM package (Hsu et al. 2003; Chang et Lin 2011) while Elastic Net relies on the widely-used implementation Glmnet (Friedman et al. 2010). Users may select between support vector classification (SVC), probabilistic support vector classification, support vector Regression (SVR), Elastic Net classification and Elastic Net Regression. Note that while regression is corresponds to the GLM offered in GraphVar, at this point we do not offer logistic regression, the complement to classification. Furthermore, the GraphVar ML framework allows a straightforward extension to any other regularization method by an advanced user. In Elastic Net, the trade-off between L1 and L2 regularization is controlled by the alpha parameter, while the lambda parameter controls the strength of the penalty on the coefficients. In SVC and SVR parameter C optimizes the size of the hyperplane margin. The parameters are preset for all GraphVar ML learning methods but may either be tuned as part of the nested cross validation procedure or manually overwritten by the user through the Manual Model Tuning panel.

**2.4 Model selection and validation**

On top of providing a standard K-fold cross-validation approach, GraphVar ML provides a nested cross validation option with a 3-step nested cross-validation structure for prediction model validation (Whelan et al. 2014). The nested cross-validation structure consists of an outer loop (final validation), an optional middle loop (hyperparameter optimization) and an optional inner loop (feature selection). The hyperparameter optimization and feature selection options are part of the model selection procedure. Intuitively, nested cross-validation introduces an additional data split, using part of the training data for the lower hierarchy level. Within each hierarchy level, the parameters from the winning model with the best performance are carried over to the higher level. One may think of it as a recursive process (Fig. 4).

**2.4.1 Cross-Validation – The Outer Loop:** In order to test the predictive performance of a trained prediction model, the model needs to be evaluated on new, unseen data in a process known as validation. A common approach in machine learning is to hold out part of the available data as a test set. By doing so, one obtains an estimate of the expected out-of-sample model performance. However, such partitioning reduces the sample size of the training data, increasing the risk of 1) losing important patterns in the data set, 2) introducing a result dependency on a particular random choice for a pair of train and test sets.

K-fold cross-validation is a commonly used approach to overcome these limitations, partitioning the overall sample in a way that allows independence whilst employing the same available data. In K-fold cross validation, the data is divided into K subsets, repeating the holdout method K times, such that each time, one of the K subsets is used as the test set/validation set and the other K-1 subsets are combined to form a training set. The resulting predictions are then averaged over all K trials to assess the overall performance of the model. The choice of K in K-fold cross validation is debatable. While many algorithms conventionally use a K number of 5 or 10, the selection is subject to a bias-variance trade-off. A lower K is computationally cheaper, produces less variance and results in more bias, while a higher K is more expensive, produces more variance and a lower bias (Kohavi 1995). GraphVar ML allows users to manually select the number of K folds. If nested cross-validation is selected, the K fold number in the outer loop is carried over to the middle and the inner loop.

Notably feature scaling within GraphVar ML is implemented within each individual test and train split, ensuring that the scale of the training set is applied to the test set within each fold. This prevents data leakage, where the predictive model is unintentionally informed by the unseen, i.e. holdout data, invalidating the estimated prediction performance (Kaufman et al. 2012).

**2.4.2 Nested cross-validation – Hyperparameter optimization in the Middle Loop:**

Hyperparameters determine how the model is fit, for example how much regularization is applied. Hyperparameters are typically predefined prior to model training, unless they are optimized through model tuning or selection, i.e. in a nested cross-validation approach (Kristajic et al. 2010; cf. Varoquaux et al. 2017). Hyper- parameter optimization is typically performed using grid search whereby a model is trained and evaluated across a combination of possible hyperparameters. The goal of such a grid search procedure is to determine a winning combination of hyperparameter values that yields the best predictive performance when evaluated by cross-validation.

In GraphVar ML, the exhaustiveness of the grid search may be customized by selecting the number of optimization steps, i.e. N. The size of the parameter space is directly determined by this choice (cf. Hastie 2001). A higher number of steps may be computationally expensive, due to high dimensionality. The range of the hyperparameters "alpha" and "lambda" in Elastic Net as well as "C" in SV are predefined within GraphVar ML. The number of hyperparameter optimization steps determines the scaling of the individual step size (see Table 1).

**2.4.3 Nested cross-validation - Feature selection in the Innermost Loop:**

Typically, the goal of feature selection is to improve predictive performance, provide faster, more cost effective predictors and allow a better intuition of the underlying process that generated the data (Guyon 2003).

On top of any feature selection which may be performed during model regularization, GraphVar ML can perform additional optional feature selection employing a relative user-selected threshold. Feature selection is implemented by ranking feature weights at the innermost loop. A relationship between the individual features in the design matrix and the predicted outcome at the corresponding nested fold is determined, thereby estimating the contribution of each feature to the model. Here, the relationship is calculated using a linear model. This implementation assumes that the individual features are not correlated - as feature overlap cannot be accounted for.

**2.5 Model performance - parametric and non-parametric testing**

Regression and classification models in machine learning are typically evaluated using a set of standard performance measures, or metrics. Metrics derived through parametric testing alone describe discrepancies between known (i.e. actual) and predicted values. Non-parametric, i.e. permutation, assessment allows an evaluation in a null-distribution, in order to assess if a real label structure, that is, a real connection between the data and the labels was established by the trained model (Ojola et al. 2010).

**3.0 Methods: Model Exploration and Evaluation**

GraphVar offers an interactive viewer that allows intuitive exploration of model performance. Figure 5 depicts available GraphVar ML plots as a function of the selected outcome variable type (i.e. classification or regression). Users may individually assess results across selected outcome variables. The interface also allows performance comparison at multiple network thresholds. Furthermore, the p-value based significance of individual feature weights in each model may be viewed along with the corresponding feature name. If a nuisance covariate is included as a feature, an additional nuisance only model is built in parallel to the full model (incl. the nuisance covariate). This allows a side by side comparison between the full and nuisance only models to assess potential influences of the nuisance variable(s) on the full model. Users may export relevant result data in spreadsheet format (.csv and .xlxs). Live figure export is possible in standard formats such as .png as well as vector format quality (.eps, .pdf).

**3.1 Classification Metrics**

Commonly reported classification performance metrics are implemented (Fig. 6), allowing the user to interact with the results and assess model performance. Plots available (Fig. 5) for classification performance assessment include: (1) confusion matrix, (2) receiver operating curve, (3) precision - recall curve, (4) feature weights, (5) null distribution histogram.

The confusion matrix plot (Fig. 7) displays model performance summed over all folds. For binary classification, i.e. patients vs. controls, true positives (TP), false positives (FP), false negatives (FN) and true negatives (TN) are displayed, along with the corresponding percentage scores. The core classification performance metrics are in turn derived from the confusion matrix.

The receiver operating characteristic (ROC) curve (Fig. 8) (Fawcett 2006) is used to compare the sensitivity, or true positive rate (TPR), against the specificity, or false positive rate (FPR). Each point on the ROC curve represents a TPR/FPR pair corresponding to a particular decision threshold. In GraphVar ML, the number of decision thresholds is equal to the number of samples. The ROC curve is calculated once across all cross validation folds by combining the predictions from the different folds. Another approach may be to calculate ROC curves separately for each fold, and then to average them. This approach may be problematic in the case of smaller validation sets, where the individual per-fold ROC curves would have only very few decision thresholds, thus limiting their resolution and interpretability. Note that this approach is only valid if the predictions that are being combined have the same scale. For GraphVar ML, this is the case because we scale features before training, thus scaling the model as well.

Area under the curve (AUC) scores are also displayed. AUC scores range between 0 and 1.0, reflecting the probability that the classifier will rank a randomly chosen positive instance higher than a randomly chosen negative instance. Here an AUC score below 0.5 indicates the classifier performs below chance.

The precision-recall curve shows the trade-off between precision (TP/ TP + FP) and recall (TP/ TP + FN) for different thresholds. Here, a large AUC score represents both high recall and high precision, where high precision relates to a low false positive rate, and high recall relates to a low false negative rate. Additional performance metrics such as accuracy, error, F1 score and Matthews correlation coefficient (MCC) are also reported (see Appendix I for quick reference glossary of terms).

### 3.2 Regression Metrics

Regression performance metrics are implemented (Fig. 6) for regression model assessment. Plots available (Fig. 5) for regression performance assessment include: (1) scatter plot (2) residuals plot (3) feature weights (4) null distribution histogram. The scatter plot (Fig. 9) displays the actual vs. predicted values, along with a line of best fit and a coefficient of determination score (R2). A residuals plot allows the user to access the distribution of standardized residuals.

Additional performance metrics such as relative absolute error (RAE), root mean squared error (RMSE), normalized root mean squared error (NRMSE), relative squared error (RSE) and mean absolute error (MAE) are available to the user (see Appendix I for quick reference glossary of terms).

### 3.3 Parametric p-values and Non-parametric performance metric

Both classification and regression models may be evaluated parametrically and non-parametrically. Figure 6 depicts the performance metrics available in GraphVar ML as a function of parametric or non-parametric performance testing.

Parametric p-values are calculated for the ROC AUC in classification models as well as for feature weights for both classification and regression. Permutation based, i.e. non-parametric performance metrics (i.e. AUC, accuracy or error for classification and R2 for regression) describe the predictive performance of the model in a null distribution. This equates to estimating the predictive performance of the built model in the absence of an informative label-data relationship in the training data. A histogram plot (Fig. 10) allows users to review these results.

Non-parametric p-values are derived by placing the actual performance metric value in the corresponding null-model distribution and determining its percentile position. More specifically, permutation based assessment allows the evaluation with respect to a corresponding null-distribution (e.g. an alpha level of <.05 would be achieved if the actual performance metric falls within the upper 95th percentile of the permutation distribution). Moreover, GraphVar ML also provides non-parametric p-values for feature weights to assess the contribution of individual feature weights in each model. The number of permutations is customizable and should be chosen based on the desired significance level.

### 3.4 Feature weights in classification and regression models

In comparison to standard GLM, i.e. encoding models, machine learning methods do not show by themselves which variables are relevant for prediction, and which are not. To get around this limitation, we developed an approach which allows for the intuitive interpretation of feature weights.

Model feature weights are estimated in a mass-univariate approach using the correlations between the individual features in the design matrix and the predictions of the models at the outermost fold, with a mean score taken across all outer folds. This score estimates the contribution of each feature to the model on average (Fig. 11). This approach is based on the relationship between encoding and decoding models outlined by Haufe et al. (2014). P-values describing the significance of the feature weights are generated for a parametric as well as non-parametric distribution where appropriate. Alpha values may be chosen manually, depending on the desired significance threshold. Finally, false discovery rate (FDR) and Bonferroni method are available, allowing correction in case of alpha inflation.

### 4.0 Limitations and caveats

While measures used to evaluate model performance (e.g. accuracy, AUC, etc.) are calculated across the sum of cross- validation folds, the averaging of feature weights across folds may cause issues for the interpretability of the final feature weights. Thus, although parametric p-values for feature weights are available, we strongly encourage users to base conclusions on results derived by permutation based significance testing for the overall model feature weights. Future work may improve the sensitivity of multiple comparison correction by applying a variant of network based statistics (NBS, Zalesky et al., 2010) or spatial pairwise clustering to feature weights.

The effect of pre-processing steps may have on the data in general and in turn the interpretability of the decoding model performance, remains an open question outside the scope of this paper. It should be noted that the GraphVar toolbox primarily targets fMRI resting state analyses. Other modalities, such as MEG and EEG may require additional pipeline complexity. GraphVar cannot account for pre-processing induced variability (please refer to Appendix II, where a short summary on the issue of critical rs-fMRI preprocessing and network construction choice is provided). However, the toolbox allows easy replication of analysis pipelines for different choices of pre-processing, which may aid and encourage investigations into these issues.

Of note, GraphVar currently focusses on a limited number of methods for the construction of networks (c.f., Kruschwitz et al., 2015) and second-level statistical analyses (i.e., GLM and machine-learning) and does not claim to provide an exhaustive collection of statistical connectomics. Methods currently not implemented but also highly relevant may include among others: multivariate distance matrix regression (Shehzad et al 2014), sum of powered score (Kim et al 2014), graphical lasso (Friedmann et al, 2007), network-constrained lasso (Li and Li, 2008; Azencott et al, 2013). For a more complete review on statistical connectomics please refer to Chapter 11 in Fornito, Zalesky et Bullmore (2016).

## 5.0 Conclusion

Here we introduced GraphVar ML, the new machine learning extension to the GraphVar GUI-based MATLAB toolbox www.nitrc.org/projects/graphvar or www.rfmri.org/GraphVar. GraphVar ML allows users to easily build, deploy, and evaluate decoding models built using a comprehensive range of functional connectivity measures as well as additional subject data. Asides from introducing the key concepts involved in building and evaluating a decoding model we provide an overview of the newly available functionalities. We anticipate that the new version of GraphVar makes machine learning more accessible to a wider audience of neuroimaging researchers and clinicians and may especially encourage reproducibility in resting state decoding models.

Besides from building and assessing decoding models for precision psychiatry, GraphVar ML provides a conceptual framework for the exploration of machine learning models including network topological measures, allowing for data driven network based decoding outside of neuroimaging applications. However, such execution may likely involve additional customization and its discussion is beyond the scope of this paper. Furthermore, the implemented GUI- controlled, customizable cross-validation model construction and evaluation framework may be used as an alternative to the inbuilt MATLAB machine learning functionalities.

## 6.0 Conflict of Interest

We have no conflict of interest to declare.

**7.0 Acknowledgement**

We would like to thank Dr. Stefan Haufe of the TU Berlin for his helpful comments during our meeting. This work was supported by the German Research Foundation (SFB940/2 2017) to Technische Universität Dresden.

# Figures:

**Fig. 1:** Schematic of the general GraphVar workflow including the new GraphVar ML extension. The GraphVar graphical user interface (GUI) allows comprehensive customization of the analysis pipeline for the data generation as well as machine learning steps. A separate, interactive results viewer allows the user to selectively explore the model results after parametric or permutation-based (non-parametric) testing.

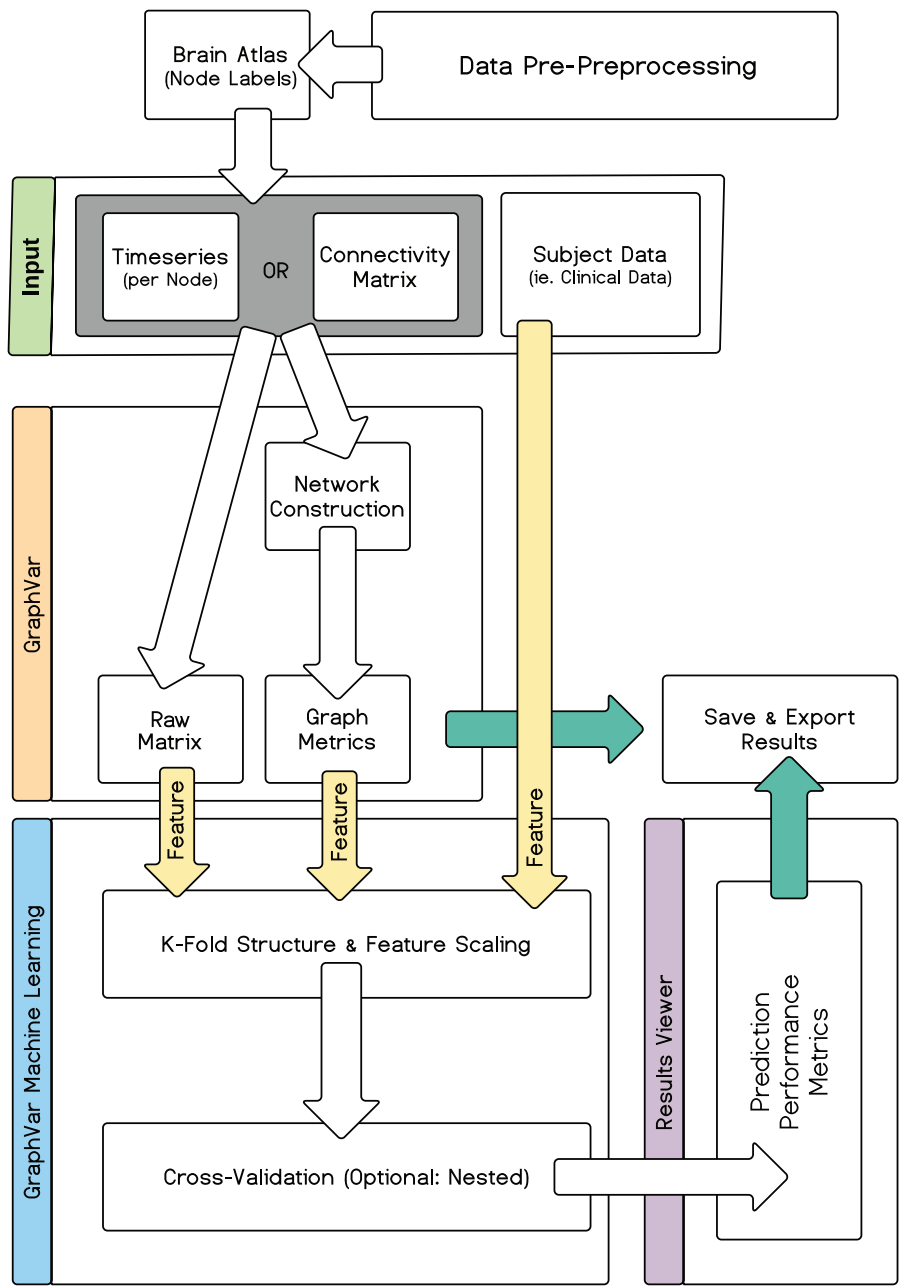

**Fig. 2:** Screenshot of the GraphVar ML extension within the GraphVar GUI. Users may select between building classification or regression models using either SV or Elastic Net. Users may also choose between executing simple cross-validation, nested cross validation (recommended) or entering hyperparameter(s) manually (i.e. for reproducibility testing). Additional (i.e. for external variable-only models) variables may be added as features through the additional variables field. The nuisance covariates field allows users to add nuisance covariates to the model, evaluating their effect by comparing the full model (incl. nuisance) against the nuisance-only model executed in parallel. Model performance may be evaluated using parametric or permutation testing. The number of permutations may be entered manually. For convenience, multiple outcome variables may be selected, cued and executed consecutively and will appear next to each other in the live results viewer.

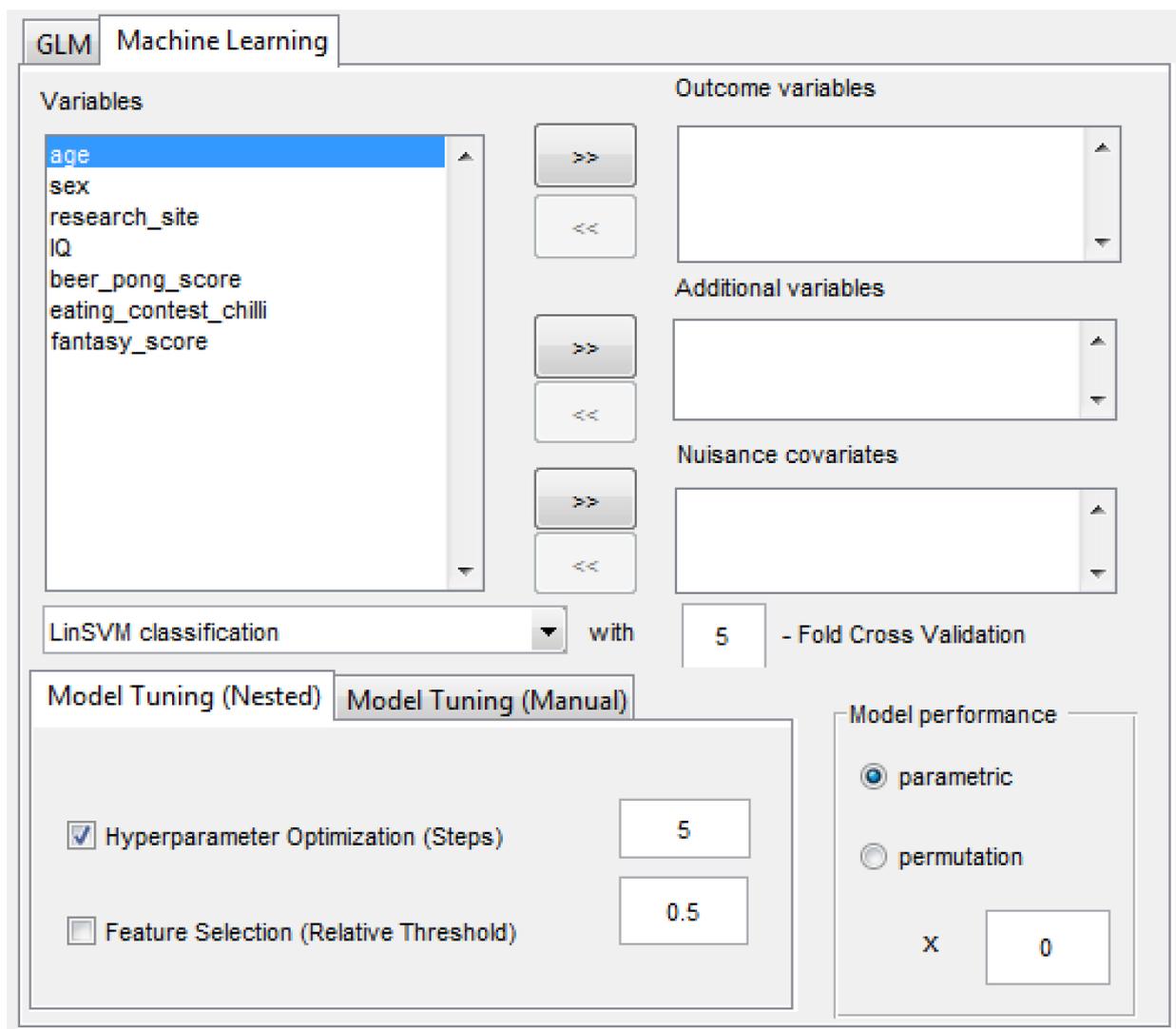

**Fig. 3:** Combining features into a design matrix. Users are free to choose any feature combination when populating the design matrix. The term graph metric and graph measure are used interchangeably. Seven cases are possible: (1) graph measure(s) alone, (2) raw connectivity matrix alone, (3) standalone additional variable, (4) graph measure and raw connectivity matrix, (5) graph measure and additional variable, (6) raw connectivity matrix and additional variable, (7) combination of all three data types. Optionally, features may be generated in dynamic connectivity mode using summary metrics.

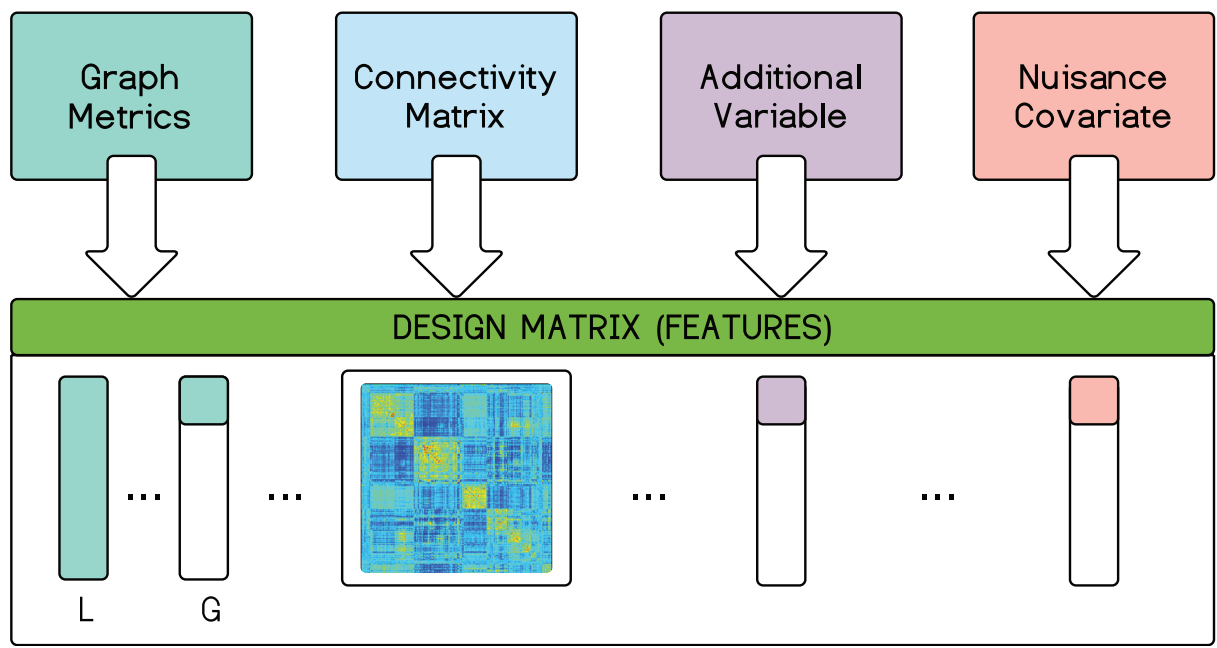

**Fig. 4:** Cross-validation structure implemented in GraphVar ML. The GraphVar ML panel allows for a custom entry of K for the outermost loop (final validation), which is automatically adapted for the middle and inner loops. If hyperparameter optimization is selected, GraphVar ML executes optimization using a grid search, with N steps, accordingly (middle loop). If feature thresholding is selected GraphVar performs feature selection inside the innermost loop.

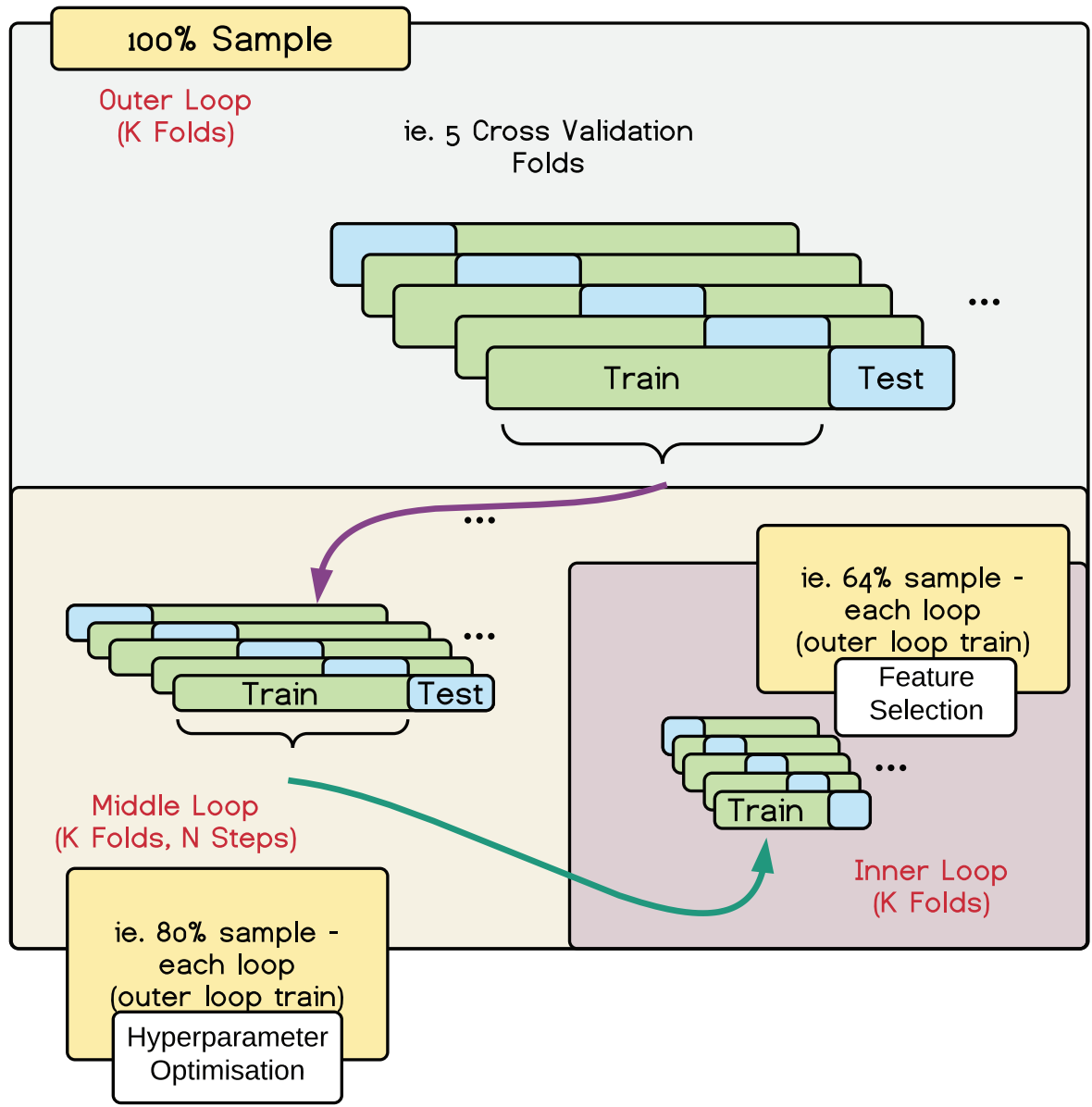

**Fig. 5:** Results plots options for GraphVar ML. The interactive GraphVar ML results viewer generates standard plots for regression and classification model performance evaluation. P-values for both parametric and non-parametric distributions are available where relevant (shaded plots).

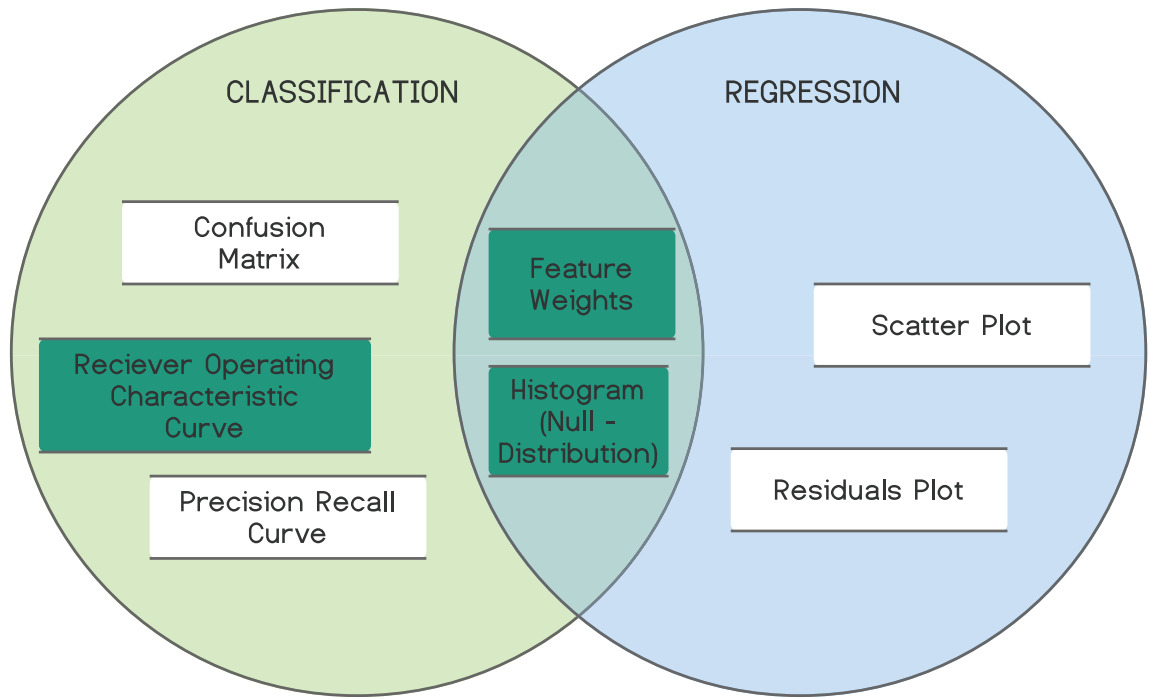

**Fig. 6:** An overview of the model performance metrics available in GraphVar ML. Both classification and regression models may be evaluated using parametric and non-parametric tests. Classifier and regressor performance metrics provide an overview of the prediction performance while feature weights provide a possible interpretation to feature significance specific to each model.

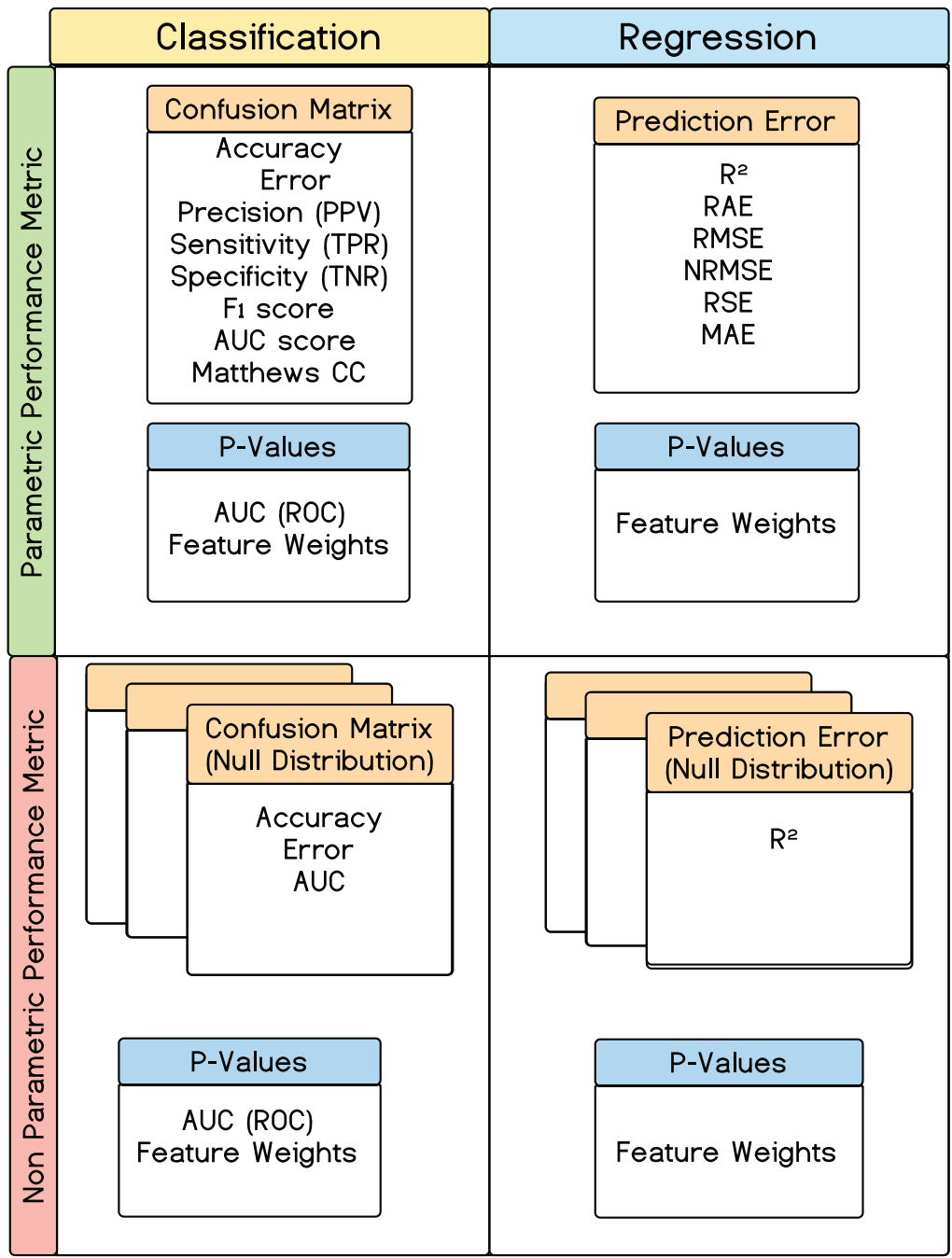

**Fig. 7:** Confusion matrix plot inside GraphVar ML Result Viewer. Confusion matrix for binary classification. True Positives (TP) i.e. hits, and True Negatives (TN), i.e. correct rejections, appear green while False Positives (FP) i.e. Type 1 error, and False Negatives (FN), i.e. Type 2 error, appear red. Overall percentage scores are displayed underneath number counts.

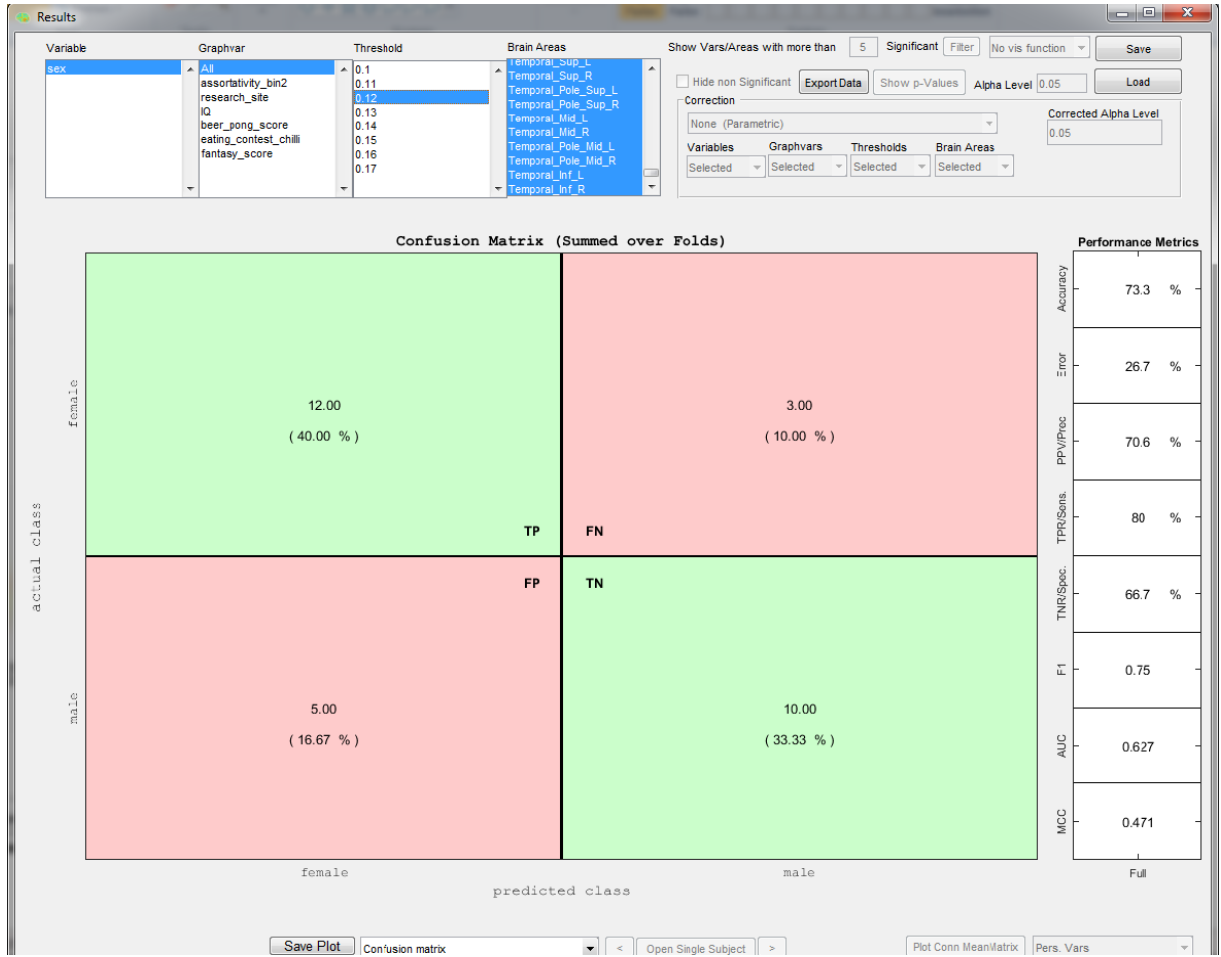

**Fig. 8:** ROC plot inside GraphVar ML Result Viewer. Sample ROC curve for binary classification. In GraphVar ML, each point on the ROC represents a sample. A diagonal line indicates the 0.5 AUC threshold. An ROC curve under the threshold line indicates chance performance.

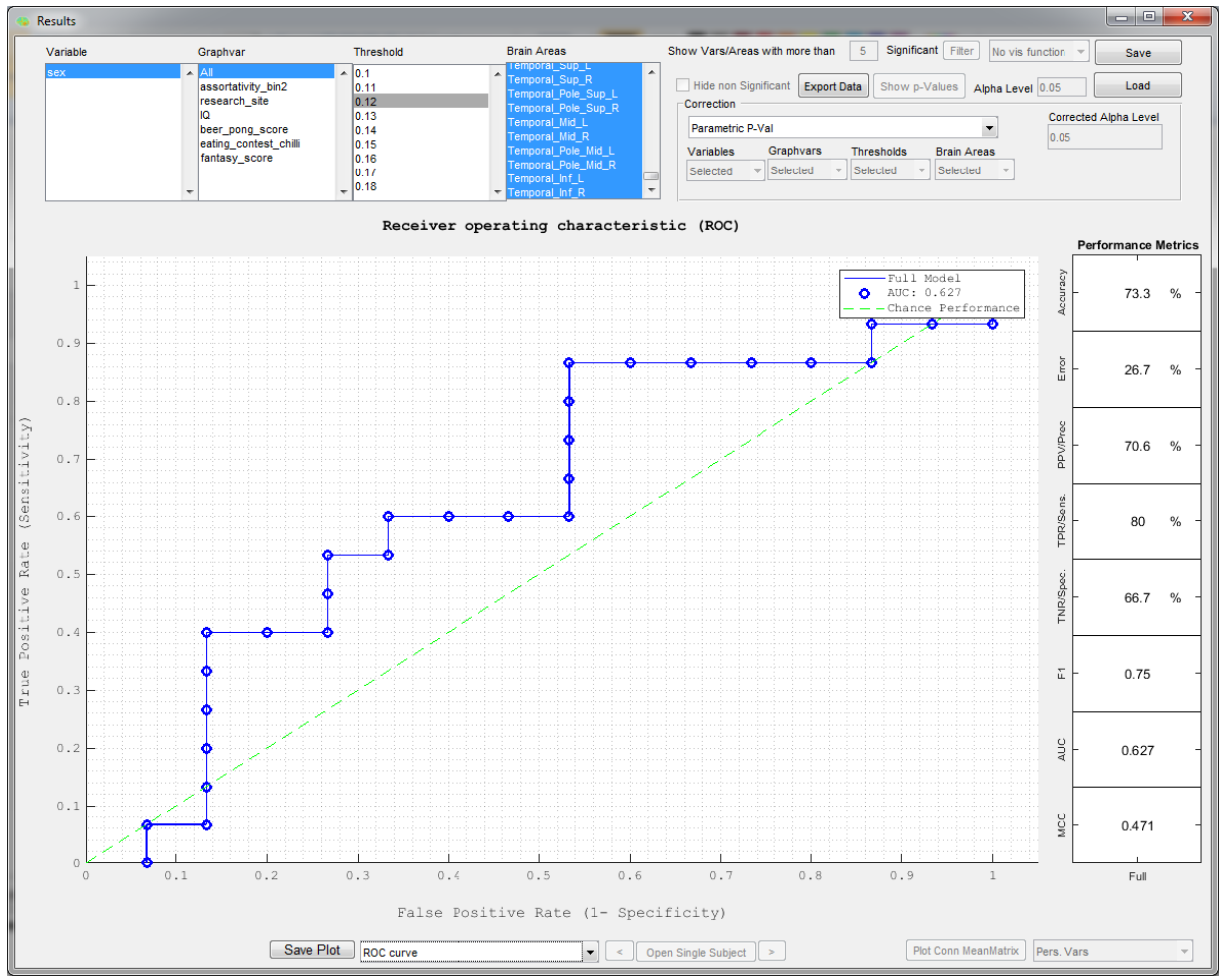

**Fig. 9:** Scatter plot inside GraphVar ML Result Viewer. Example results of a prediction with Elastic Net (regression). The full (blue) as well as nuisance-only (red) model may be compared simultaneously.

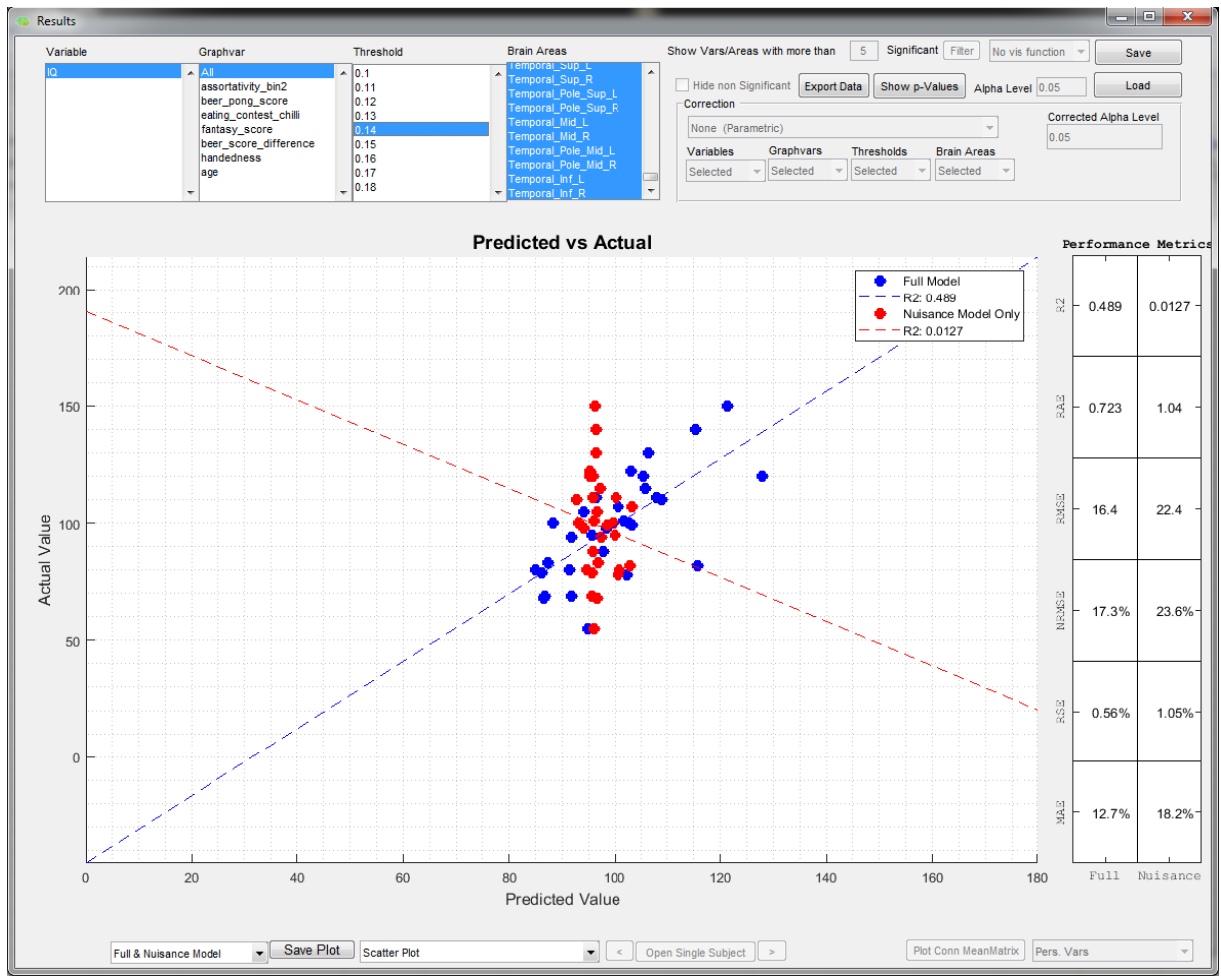

**Fig. 10:** Histogram plot inside the GraphVar ML Results Viewer. The option is available for both, regression and classification. The performance metric for the actual model can be compared against the null distribution. Here, the performance of a classifier is assessed with the AUC, comparing full model (blue) and a nuisance model distribution (red) are displayed for simultaneous comparison.

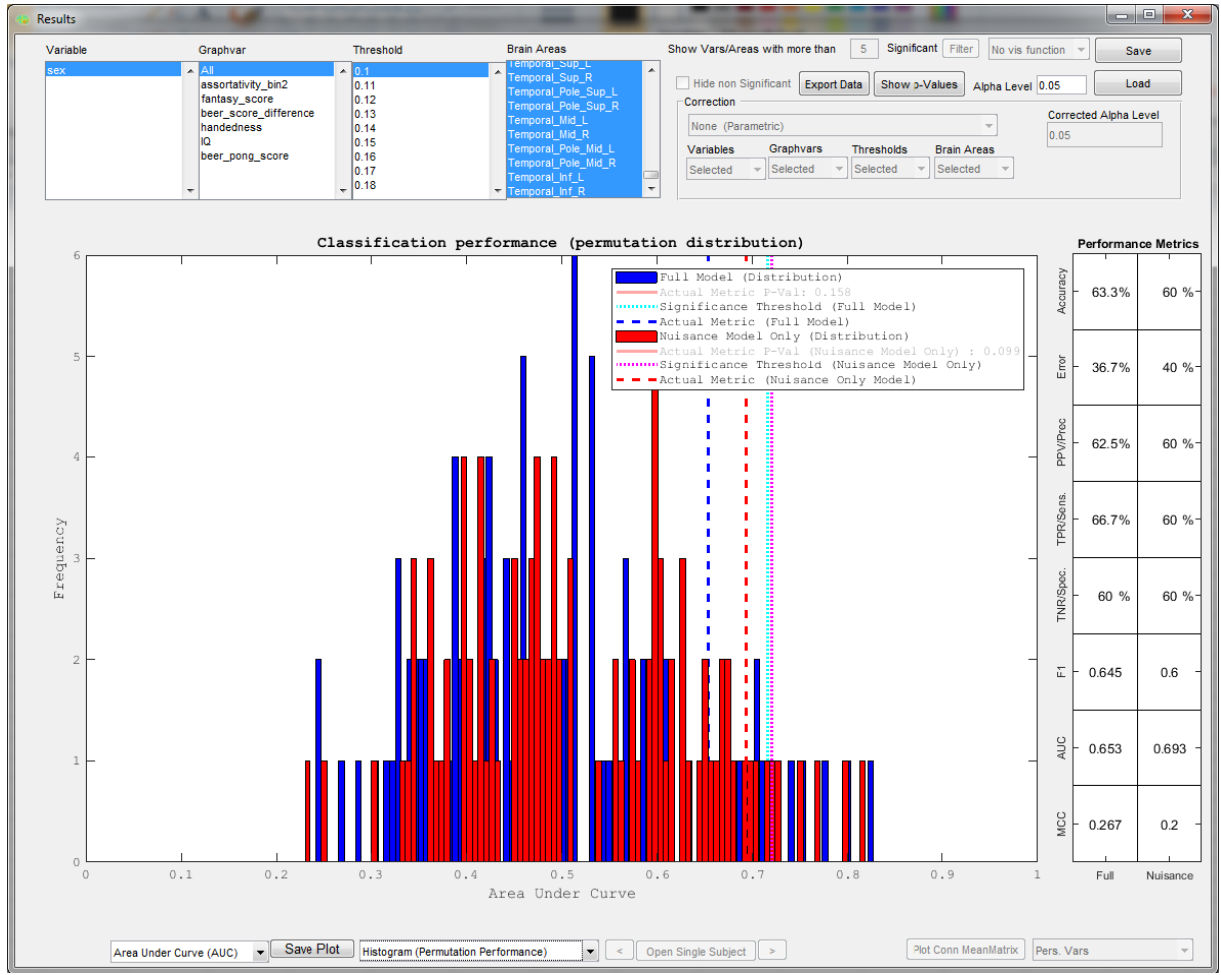

**Fig. 11:** Feature Weights plot inside the GraphVar MLResults Viewer. The weight scores and corresponding p-values of a connectivity matrix are selected for a review. Features may be inspected individually for their corresponding contribution to the overall model.

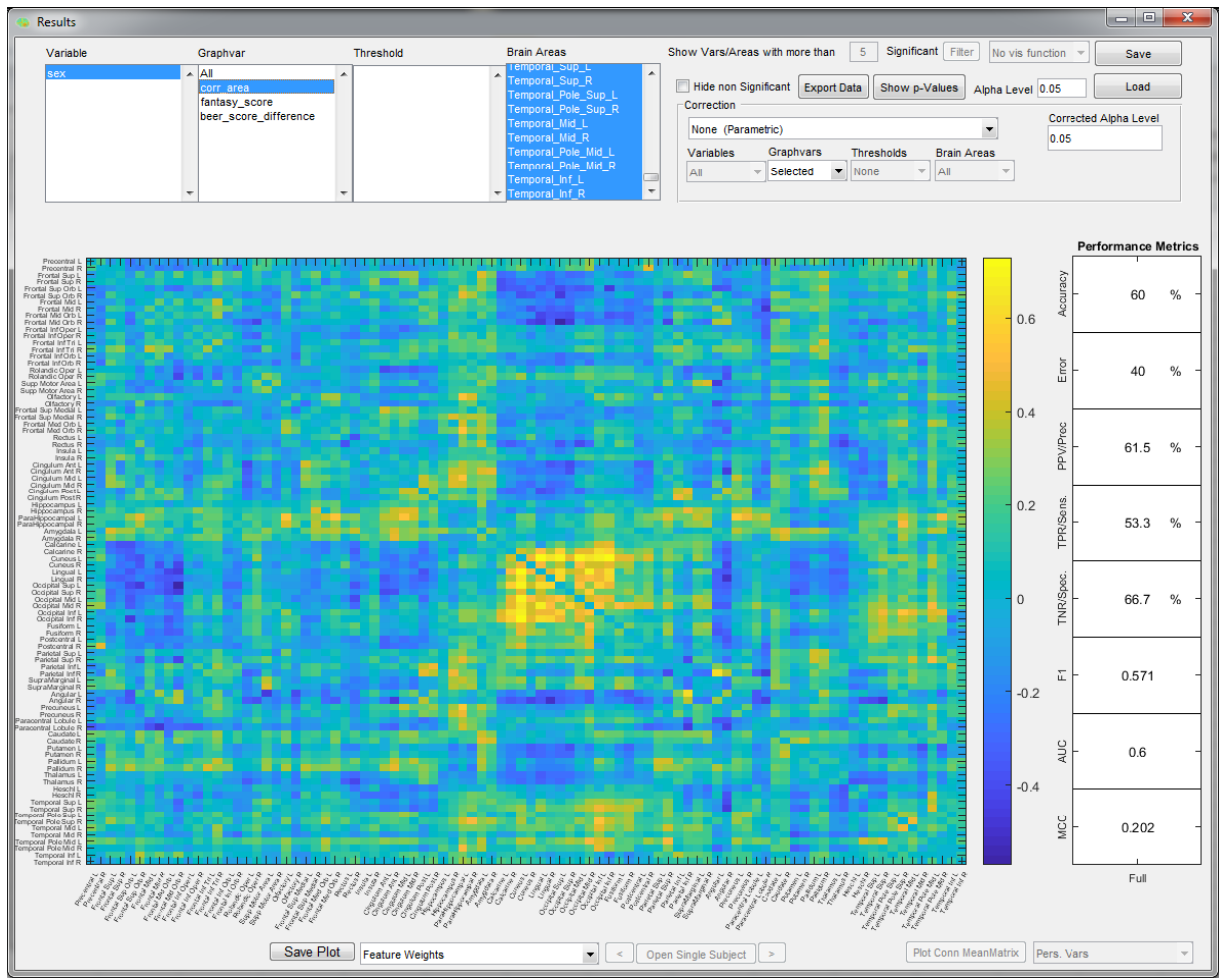

**Tables:**

**Table 1:** Grid search details for GraphVar ML. Hyper-parameters associated with each available learning model, along with their respective spacing scales and possible ranges. Steps are even and determined by the choice of N. (*cf. Hsu et al. 2003).

| Learning model | Details | | |
|---|---|---|---|
| | **Parameter** | **Spacing scale** | **Range** |
| Linear SV classification (SVC) | C | logarithmic | $10^{-2}$ to $10^{3}$ |
| Linear SV regression (SVR) | Nu | linear | 0 to 1 |
| Elastic Net classification | Alpha | linear | 0 to 1 |
| | Lambda | logarithmic | $10^{-2}$ to $10^{3}$ |
| Elastic Net regression | Alpha | linear | 0 to 1 |
| | Lambda | logarithmic | $10^{-2}$ to $10^{3}$ |

# Appendix I

**Table 1.** Quick reference glossary of terms in GraphVar ML.

| Term | Description |
|---|---|
| **Encoding** | i.e., Forward model. (such as GLM) How could data be generated? |
| **Decoding** | i.e., Backward model. Could one extract/identify informative features? |
| **Regression** | Predict a continuous quantity. e.g. "age" |
| **Classification** | Predict a discrete class or group label. e.g. "control". |
| **Binary Classification** | Determine membership of 2 groups. e.g. "patient" vs. "control". |
| **Feature** | Individual property of data. e.g. a chosen measure. |
| **Design Matrix** | i.e., Feature vector. Initially selected features to be considered in model training. |
| **Feature scaling** | Correct value range for features, decrease possible wide value range (variance). |
| **Standardization** | Type of feature scaling. Gives features properties of a standard normal distribution. |
| **Data leakage** | Properties/information from the validation data is used during model training. |
| **Overfitting (Model)** | Model fails to generalize well when validating on unseen data. |
| **Regularization** | Tuning the complexity of a model to ensure it does not overfit. |
| **Sparse model** | Model with small number of nonzero parameters or weights, opp. To "dense" model. |
| **L1 (Regularization)** | In-build feature selection. Output is sparse. |
| **L2 (Regularization** | Has analytical solution. No feature selection. Output is non-sparse. |
| **Support Vector** | Points that define i.e., "support" the decision boundary i.e., hyperplane. |
| **Elastic Net** | Linearly combines L1 and L2 regularization in Elastic Net regularization. |
| **Alpha (Parameter)** | Sets the ratio between L1 and L2 regularization in Elastic Net regularization. |
| **Lambda (Parameter)** | Sets the strength of the penalty on the coefficients in Elastic Net regularization. |
| **C (Parameter)** | Penalty parameter of error term (in linear SVM). Separating hyperplane margin = small (large C) or large (small C). |
| **Model selection** | i.e., model tuning. In nested cross-validation, selection of wining model settings. |
| **Hyperplane optimization** | Determine optimum choice for tunable model parameters (i.e. penalty terms) across a custom range. (e.g. nested cross-validation) |
| **Grid search** | Exhaustive search through a manually specified subset of the hyperparameter space. |
| **Feature selection** | Selection of subset of relevant features for use in model construction. |
| **Feature weights** | Weight attributed to individual features with regard to model i.e., prediction performance. |
| **Nuisance covariate** | i.e., nuisance variable. e.g., gender or age. which is not of interest in the analysis. |
| **Accuracy** | Ratio. Correctly predicted observation to total observations. **Best use:** balanced (class) sample datasets. **Avoid:** 1class dominates sample. |
| **Error** | Inverse of accuracy. Sometimes reported alt. to accuracy. |
| **Precision** | Ratio. **Intuition**: When model predicts condition positive, how often is it correct? |
| **Sensitivity (i.e., Recall)** | **Intuition**: When condition is actually positive, how often does it predict con. positive? |
| **Specificity** | **Intuition**: When condition is actually negative, how often does it predict con. negative? |
| **F1 (Score)** | Harmonic average of precision and recall. |
| **AUC (Score)** | Derived from ROC curve. Ranges from 1 (perfect) to zero. 0.5 indicates performance at random. |
| **Matthews CC** | Balanced measure for unevenly class balance in binary classification. Ranges from 1 (perfect) to -1. |
| **Standardized residuals** | Ratio. Plotting visualizes residual dispersion patterns on standardized scale (detect potential outliers). |
| **R2** | Coefficient of determination. Ranges from 0 to 1 (perfect). |
| **RAE** | Relative absolute error. Allows comparison between models whose errors are measured in different units. |
| **RMSE** | Root-mean-square error/deviation. Average absolute error. Lower = better fit. |
| **NRMSE** | Normalized RSME. Allows comparison between datasets or models with different scales. |
| **RSE** | Residual standard error. Large values indicate a poor fit. |
| **MAE** | Mean absolute error. Indicates magnitude of the error. Does not specify if over or under (prediction). |

**Appendix II**

*Short summary on the issue of critical rs-fMRI preprocessing and network construction choice*

Although the effect of pre-processing strategies on rs-fMRI data in general, and specifically on network construction and graph theoretic network measures, remains an open question that is outside the scope of this paper, we want to provide a short summary about this critical issue with the aim to direct the users` awareness to the importance of correct preprocessing and network construction.

Rs-fMRI data is rich in confounds and noise. For this reason, it requires adequate preprocessing, and the choice of preprocessing steps may greatly impact reliability (c.f. Murphy et al. 2013, Bright et Murphy 2015, Parkes et al 2017). Although some all-in-one preprocessing tools have emerged (e.g., DPABI, Yan et al. 2016; FMRIPrep, Esteban et al. 2018) in parallel to major fMRI analysis toolboxes (e.g., CONN, https://sites.google.com/view/conn/; SPM, www.fil.ion.ucl.ac.uk/spm/; FSL, https//.fsl.fmrib.ox.ac.uk), there is currently no gold standard on how to definitely deal with the largely heterogeneous noise. Moreover, it has been shown that the choice and the ordering of preprocessing steps may impact network topological graph measures (e.g., Braun et al., 2012; Andellini et al., 2015; Aurich et al., 2015; Gargouri et al., 2018). Due to this complexity, careful considerations should be given on how to preprocess the data for most adequate use with respect to the research hypothesis.

To guide the user, we recommend two recent studies (Ciric et al., 2017; Parkes et al., 2018) that highlight and evaluate various motion correction strategies and pipelines for rs-fMRI data, as well as a review (Andellini et al., 2015) highlighting the impact of various preprocessing steps on graph measures. Specific thoughts may be given on whether to regress out the global mean signal (GSR) or not. Regression against the global mean signal has been shown to shift correlation distributions from a largely positive mean towards a mean of zero by artificially introducing negative correlations. This is a fundamental argument against application of GSR (Murphy et al., 2009; Weissenbacher et al., 2009, as many graph theoretic measures cannot deal with negative edge weights, so these edges may need to be discarded for subsequent network analyses (Rubinov and Sporns, 2010). On the other hand, and in favor of GSR, the global signal is highly associated with motion and respiratory artefacts (Power et al., 2017). Furthermore, it has been argued that GSR may also remove a true shared covariation in neuronal firing (i.e., a true global neuronal signal), thereby revealing the true connectivity of neuronal populations that would have otherwise been masked by the dominant global signal (Fox et al., 2009; Schölvinck et al., 2010; Keller et al., 2013; Kruschwitz et al., 2015).

Of fundamental importance for reliable assumptions about the underlying data are also choices of network properties such as nodes and links. As nodes and links define a graph (the basis for network topological metrics), the neurobiological interpretation is constrained by this choice. While network nodes can range from the voxel-level to a broader macro-scale, it is important that nodes should have intrinsic consistency, extrinsic differentiation, and be spatially constrained (Fornito, Zalesky & Bullmore, 2016). Studies that compared strategies of node selection (i.e., parcellation scale) found whole-brain network topological measures to be robust over different scales but highlighted that node specific parameters can vary with the size of the parcellation scale (Fornito et al., 2010; Hayasaka and Laurienti, 2010). Despite appropriate correction of confounds such as physiological noise (with or without GSR), thoughts should also be given in how to define network links. These can range from decisions about the frequency range that will be considered after band-pass filtering (where wider ranges have been shown to result in more reliable graph metrics; e.g., Braun et al., 2012; Andellini et al., 2015) to the choice of the estimation of a measure of connectivity between nodes and definition of a significance threshold for selection of substantially contributing links (e.g. Zalesky et al., 2010). While most studies use Pearson correlation to estimate connectivity between nodes, partial correlation may be preferable for some network metrics (Rubinov and Sporns, 2010; Kruschwitz et al., 2018). It is also noteworthy that Smith et al. (2011) found methods based on covariance to be quite sensitive to the underlying network, as well as several Bayesian net methods.

*Appendix II – References*